\documentclass{PoS}

\title{Color superconductivity and dense quark matter}

\ShortTitle{Color superconductivity and dense quark matter}

\author{\speaker{Massimo Mannarelli}\thanks{This work has been supported by the Spanish grant
AYA 2005-08013-C03-02.}\\\
        Instituto de Ciencias del Espacio (IEEC/CSIC) Campus Universitat Aut\`onoma de Barcelona, Facultat de Ci\`encies, Torre C5, E-08193 Bellaterra (Barcelona), Spain\\
        E-mail: \email{massimo@ieec.uab.es}}

\abstract{The properties of cold and dense quark matter  have been the subject of extensive investigation, especially in the last decade.  Unfortunately, we still lack of a complete understanding of the  properties of matter in these conditions. One possibility is that  quark matter is  in a color superconducting phase which is characterized by the formation of a diquark condensate.   We  review  some of the  basic concepts of color superconductivity  and some of the  aspects of this phase of matter which are relevant for  compact stars. Since quarks have color, flavor as well as spin degrees of freedom many different color superconducting phases can be realized. At asymptotic densities QCD predicts that the color flavor locked phase is favored. At lower densities where the QCD coupling constant is large, perturbative methods cannot be applied and one has to rely on some effective model, eventually trying to constrain such a model with experimental observations. The picture is complicated by the requirement that matter  in the interior of compact stars is in weak equilibrium and 
neutral. These conditions and  the (possible) large value of the strange quark mass conspire to separate the Fermi momenta of quarks with different flavors,  rendering  homogenous superconducting phases unstable.  
One of the aims of this presentation is to introduce non-experts in the field to some of the basic ideas of color superconductivity and to some of its open problems.}

\FullConference{8th Conference Quark Confinement and the Hadron Spectrum \\
         September 1-6 2008\\
         Mainz, Germany}

\begin{document}
\section{Introduction}
Quantum Chromodynamics (QCD) is a relatively young theory and we still have to understand 
many of its  fundamental properties. We have good control on QCD at high energy scales, because of the property of  asymptotic freedom, but  we still do not have under quantitative control this theory at intermediate scales. In particular we lack of a picture for describing quark matter at arbitrary densities and temperatures. 

The possibility of having "liberated" quark matter at asymptotic densities~\cite{Collins} and Cooper pairing of quarks~\cite{Barrois}  was realized soon after the discovery of asymptotic freedom and is related to the fact that in QCD the  quark-quark interaction is strong and attractive between quarks that are antisymmetric in color. However, the first works on this subject shown that the effect of pairing was small, with a gap in the fermionic  quasiparticle spectrum of the order of 100 keV~\cite{bailin}, at most.  The breakthrough came at the end of 1997  when two independent groups \cite{Alford:1997zt, Rapp:1997zu} came to the conclusion that the fermionic gap might be as large as $100$ MeV with a  critical temperature  of the same order. This result boosted the investigation of the possible color superconducting (CSC) phases and of their properties~\cite{reviews}. 

One of the goals of the high density QCD  program is to reach a qualitative and semi-quantitative understanding of the properties of color superconducting quark matter 
and then discerning how its presence would affect astronomically observed or observable phenomena. The focus is mainly on  cold  and dense stellar objects that are generically called compact stars. However, after about four decades of investigation, it still  remains an open question whether  "liberated" quark matter can exist at the core of compact stars, which have relatively small central densities, at most $\sim 10$ times that of ordinary nuclear matter.

\begin{figure*}[t]
\label{phasediagram}
\begin{center}
\includegraphics[width=7cm,angle=0]{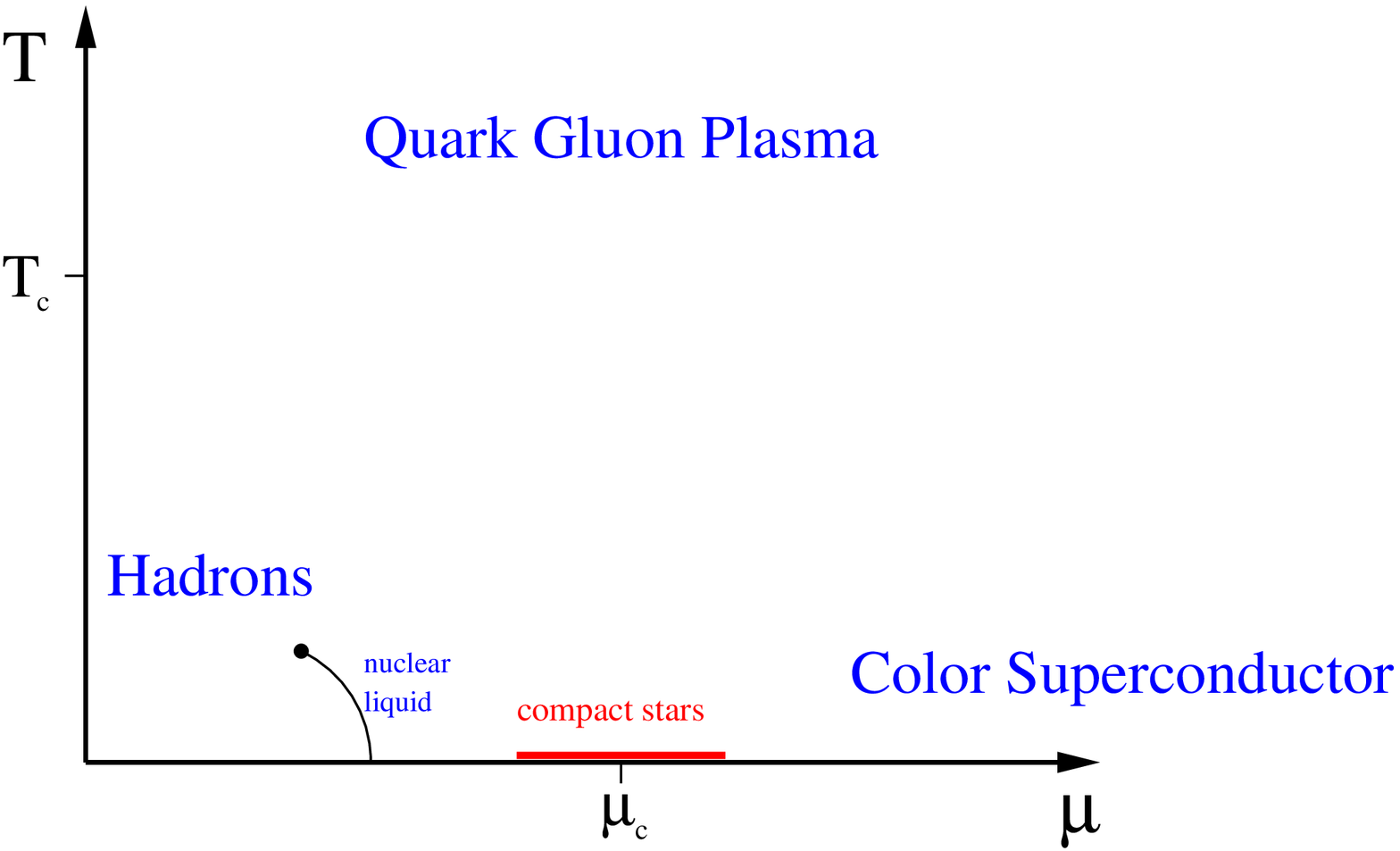}
\includegraphics[width=7cm,angle=0]{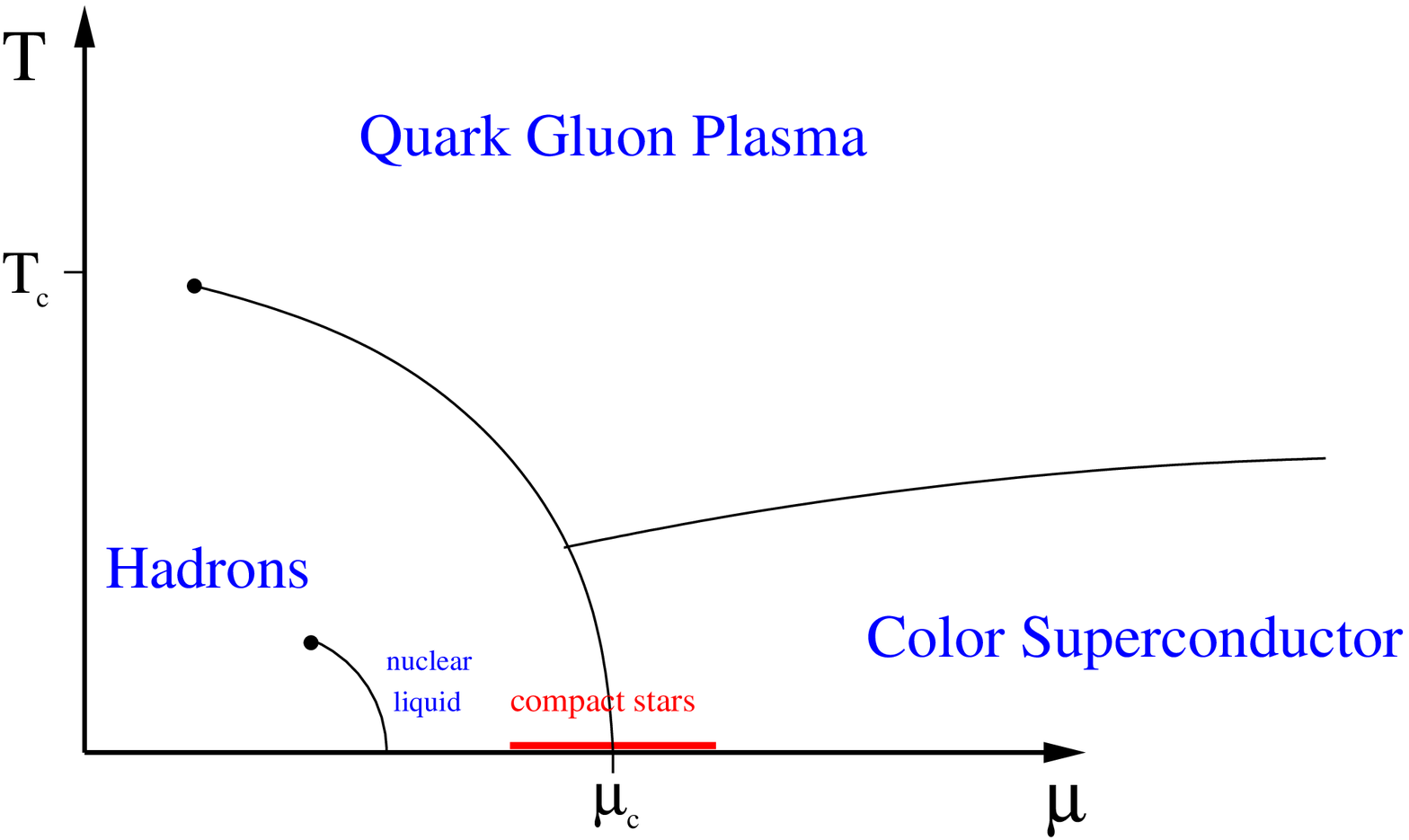}
\caption{Cartoon of the QCD phase diagram. The critical temperature $T_c \sim 200$ MeV indicates the temperature for the onset of a phase where quark and gluons are the fundamental degrees of freedom. At extremely high temperature matter forms the Quark Gluon Plasma, i.e. it can be described as a system of weakly interacting quarks and gluons. On the horizontal-axis it is reported the quark number chemical potential. The critical potential $\mu_c \sim 300-500$ MeV indicates the possible  onset of  a new phase. At extremely high chemical potentials the ground state of quark matter is the color superconducting phase. The left-hand side represents our present knowledge. The right-hand side represents one of the possible scenario, see the text for details.}
\end{center} 
\end{figure*}

Our understanding of the various phases of matter can be summarized in the so-called QCD phase diagram that is reported on the left-hand side of  Fig.~\ref{phasediagram}. This is a schematic view of the possible phases of matter as a function of temperature and quark number chemical potential (more detailed analysis of the QCD phase diagram are presented in various talks of these proceedings). 
For large energy scales we have control on the theory because the QCD coupling  is small and perturbative methods can be applied.  On the other hand, the lack of lines separating the various phases at intermediate scales reflects  our poor knowledge of QCD at these scales. 

Heavy-ion experiments and lattice simulations give us many
informations on the behavior of matter at low chemical potentials
and at high temperatures. In particular lattice simulations indicate  that at vanishing chemical potentials there is a smooth crossover between the confined and the  deconfined phase at a critical temperature $T_c \sim 200$ MeV.  On the other hand, no currently available calculational method can help us to
understand the properties of matter at large but not asymptotic
chemical potentials (lattice simulations at non-zero chemical potentials are still in a
preliminary stage). Therefore, investigating the existence of a phase transition  and eventually determining the value of $\mu_c$ where nuclear matter turns to color superconducting phase (or to some other  phase) is
not a task that  can be probably accomplished any time soon. This is not
surprising since locating such a phase transition requires a
quantitative detailed control of the pressure of both phases.

With our present knowledge of QCD we can say that various scenarios may be realized. The most popular one is that there is a first order phase transition between the hadronic phase and the color superconducting  phase. This can be naively understood assuming that there is a competition between the chiral condensate and the diquark condensate. At higher temperatures the first order phase transition line should end in a critical endpoint at some non-zero chemical potential.  This possibility is sketched on the right hand side of Fig.~\ref{phasediagram}.
Alternatively there can be a hadron-quark continuity at small temperatures, meaning that there is a smooth crossover between the hadronic phase  and the CSC phase \cite{Schafer:1998ef} with the possible presence of a second critical point \cite{Hatsuda:2006ps}. Other possibilities are the presence, at least in a certain range of densities around $\mu_c$, of the quarkyonic phase~\cite{McLerran:2007qj} or the chiral density wave~\cite{Deryagin:1992rw}, that are energetically favored over the CSC phase for large $N_c$.


The lack of  {\it ab initio} calculational methods in the range of densities at which the phase 
 transition from cold nuclear matter to quark matter occurs,  forces us to use a phenomenological approach. As reported in Fig.~\ref{phasediagram} compact stars might have densities   
such that in their interior  $\mu \sim \mu_c$. Therefore,  compact stars could be a "laboratory" for studying the transition from cold nuclear matter to quark matter.  As we have already said this is not necessarily what happens in Nature, however   we believe that it is a possibility that presently can be ruled in or out  by astrophysical observations only.   

Except during the first few seconds after their birth, the
temperature of compact stars is expected to be of the order of tens
of keV at most. Such low temperatures are much smaller than the
critical temperatures of the possible CSC phases. Therefore, if deconfined quark matter exists
in the core of compact stars it is likely to be in one of the
possible color superconducting phases.
Furthermore, since the temperature of compact stars is so  small, for many purposes the quark matter that may be found in their interior is well-approximated as having zero temperature.

There are many different observables related to compact stars that can be
studied in order to obtain useful informations. In the table below  I have reported some of them and the corresponding talk in the present conference.  
\begin{center}
\begin{tabular}{|l|c|}\hline \label{tab1}
  {\bf Equation of state} & Blaschke, Burgio \\ \hline
  {\bf Formation and dynamics of compact objects}  & Schaffner-Bielich, Drago, Yakovlev
  \\ \hline
  {\bf Cooling} & Kaminker \\ \hline
  {\bf Glitches} & Sharma, M.M. \\ \hline
  {\bf Gravitational waves} & Horowitz, Pagliara \\
  \hline
\end{tabular}
\end{center}
This table is  quite schematic and by no means exhaustive, but gives an idea of what is going on in our community. We do not have time and space to deal with any single topic which  the reader will find discussed in the pertinent talk.  We just want to note the  quite interesting fact  that the  non-detection of gravitational waves by gravitational wave  interferometric detectors~\cite{Abbott:2008fx} can be employed to  have information on the possible phases of quark matter~\cite{Lin:2007rz}.

\section{Color Superconductors}
The essence of  superconductivity is the di-fermion pairing, driven by
the BCS mechanism which operates whenever there is an attractive
interaction between fermions at a Fermi surface~\cite{BCS}. These correlated pairs of fermions (Cooper pairs) behave as bosonic particles and  condense in the lowest energy state.
This is  a quite remarkable result considering that Cooper pairs may have a correlation length   (characterizing the size of the bosonic particles) much larger than the average distance between fermions.

Quite generally the presence of a di-fermion condensate determines the spontaneous breaking of some continuous  symmetry of the Lagrangian. In case a gauge symmetry is broken the system will be a superconductor and the magnetic component of the gauge field associated with the broken generator will be expelled  (Meissner effect).  In case a global symmetry is  broken the system will have a superfluid mode. These  are not two  mutually exclusive possibilities, meaning that in  case  local and global symmetries are broken the system will be superconducting and superfluid. Upon first hearing this may sound as  a remote possibility that can be realized in exotic systems. We shall see that it is not so.

The standard example of BCS mechanism is the formation of Cooper pairs of electrons in metals.
This is a quite tenuous effect because of the presence of the  repulsive Coulomb  interaction between electrons. However, this  interaction is screened and for very small temperatures the attractive interaction mediated by phonons allows the formation of Cooper pairs. Since the phonon interaction is  weak the resulting superconductor will have a small critical temperature. 
 
The formation of Cooper pairs in cold and dense quark matter  should be a  much more robust phenomenon. First of all because the interaction that drives the formation of quark pairs is the  color interaction which is much stronger than the phonon interaction. Secondly, because  quarks of different colors and flavors can participate in the formation of Cooper pairs. The BCS mechanism is a cooperative phenomenon  and the larger the number of fermions involved in pairing, the larger is the effect.  Notice that since a diquark  cannot be a color singlet,  some color symmetries will be spontaneously broken and this is the reason this phase is called color superconductor. 

\section{Extreme Densities: the CFL phase}
It is by now well established that at asymptotic densities where perturbation theory can be applied and  the masses of up, down and strange quarks can be neglected, quark matter is in the color flavor locked (CFL) phase~\cite{reviews, Alford:1998mk}. The CFL condensate is
antisymmetric in color and flavor indices and  involves
pairing between quarks of all different colors and flavors. The quark-quark condensate is given by
\begin{equation}
\langle\psi_{i\alpha}({\bf p})C\gamma_5  \psi_{j\beta}(-{\bf p})\rangle \sim \Delta_0  \epsilon_{\alpha\beta I}\epsilon_{i j I}\,, \label{condensate} 
\end{equation}
where $i,j=1,2,3$ are flavor indices and $\alpha,\beta=1,2,3$ are
color indices and we  have suppressed the Dirac indices.  $\Delta_0$ is called the gap parameter and we will explain below how it is related with the dispersion law of quasiparticles. The quark-quark condensate   induces the symmetry breaking pattern
\begin{equation}\label{breaking}SU(3)_C\times SU(3)_L \times SU(3)_R \times U(1)_B \rightarrow SU(3)_{C+L+R} \times Z_2 \,.\end{equation}
In order to understand this symmetry breaking let us consider the more familiar chiral symmetry breaking induced  by the chiral condensate. In that case  $SU(3)_L \times SU(3)_R \rightarrow SU(3)_{V} $, meaning that the only symmetry that leaves the  vacuum invariant is a simultaneous rotation of left-handed and right-handed fields.  The breaking pattern in Eq.(\ref{breaking}) is more complicated than the chiral symmetry breaking pattern because there is simultaneous breaking of global and local symmetries. The only symmetry that leaves the vacuum invariant is a simultaneous rotation of  color and flavor indices. Notice that the color symmetry is completely broken and therefore all gluons acquire a Meissner mass.   Moreover, there are  eight Goldstone bosons associated to the breaking of flavor symmetries. For non-zero quark masses these Goldstone bosons acquire a mass and the only true massless Goldstone boson is the one associated with the   breaking of $U(1)_B$ and characterizes  the superfluid properties of the system~\cite{Son:2002zn}. 

It is interesting to investigate the possibility that  strange quark matter is absolutely stable~\cite{Witten}, {\it i.e.} it has a lower energy per baryon than nuclear matter.  In this case a  quark star   consisting of CFL quark matter  all the way to the surface may exist. 
However,  CFL quark matter is superfluid and there are no  efficient mechanism for damping r-modes oscillations~\cite{Madsen:1999ci}.   R-modes oscillations are non-radial oscillations of the star that  coupling to  gravitational radiation  lead to the spin down of the star~\cite{Andersson:2000mf} and do not allow    CFL stars to rotate at  frequencies larger than  $1$ Hz~\cite{Mannarelli}.   Less than $25\%$ of observed pulsars have rotation frequencies less than $1$ Hz; being mostly isolated pulsars, their masses are unknown.

\section{The Nambu-Jona Lasinio model}
For describing the properties of matter and to obtain semi-quantitative predictions at intermediate densities one can employ the Nambu-Jona Lasinio (NJL)  model \cite{NJL}.  This is a simple phenomenological model with no gauge fields and  a contact four fermion interaction that mimics the color interaction.   At first sight the  NJL model has almost nothing to do with QCD (it lacks all the dynamics of the gauge fields, which includes the property of confinement) however it can reproduce some aspects of QCD~\cite{Buballa:2003qv}. 

We are interested in describing  quark matter containing massless $u$ and $d$
quarks and $s$ quarks with an effective mass $M_s$. The Lagrangian
density for this system in the absence of interactions is
given by
\begin{equation}
{\cal L}_0=\bar{\psi}_{i\alpha}\,\left(i\,\partial\!\!\!\!\!
/^{\alpha \beta}_{ij} -M_{ij}^{\alpha\beta}+
\mu^{\alpha\beta}_{ij} \,\gamma_0\right)\,\psi_{\beta j}
\label{lagr1}\ \,.
\end{equation}
 The mass
matrix is given by $M_{ij}^{\alpha\beta} =\delta^{\alpha\beta}\,{\rm
diag}(0,0,M_s)_{ij} $, whereas
$\partial^{\alpha\beta}_{ij}=\delta^{\alpha\beta}\delta_{ij}\partial$ and
the quark chemical potential matrix is  given by
\begin{equation}\mu^{\alpha\beta}_{ij}=(\mu\delta_{ij}-\mu_e
Q_{ij})\delta^{\alpha\beta} + \delta_{ij} \left(\mu_3
T_3^{\alpha\beta}+\frac{2}{\sqrt 3}\mu_8 T_8^{\alpha\beta}\right) \,
, \label{mu}
\end{equation} with  $Q_{ij} = {\rm
diag}(2/3,-1/3,-1/3)_{ij} $ the quark electric-charge matrix and
$T_3$ and $T_8$ the Gell-Mann matrices in color space. $\mu$ is the
quark number chemical potential and if quark matter exists in
neutron stars, we expect $\mu$  to be in the range $350-500$ MeV. The gauge chemical potentials $\mu_e, \mu_3$ and $\mu_8$ are introduced by hand in order to enforce color and electrical neutrality. In QCD  neutrality is a result of gauge field dynamics; in NJL models one has to introduce Lagrange multipliers by hand. Then, the neutrality conditions can be written as 
\begin{equation}
\frac{\partial \Omega}{\partial \mu_e} =0 ,\hspace{.5cm} \frac{\partial \Omega}{\partial \mu_3} =0 ,\hspace{.5cm}\frac{\partial \Omega}{\partial \mu_8} =0\,,
\end{equation}
where $\Omega$ is the free energy of the system. In principle one should demand that quark matter is in a color singlet state. This is a more stringent constrain of  color neutrality, but has a rather small effect on the free energy of the system for large volumes~\cite{Amore:2001uf}. 
 
In the NJL model the interaction between quarks is described by the local interaction
\begin{equation}
{\cal{L}}_{\rm interaction}\ =\ \frac{3}{8}\,G\,
(\bar{\psi}\,\Gamma\,\psi)(\bar{\psi}\,\Gamma\,\psi) \,,
\label{interactionLagrangian}\
\end{equation}
that we add to the Lagrangian density
(\ref{lagr1}), and is usually treated  in the mean field approximation. Here,
we have suppressed the color, flavor and Dirac indices;
the full expression for the vertex factor is $\Gamma^{A\nu}_{\alpha
i,\beta j} = \gamma^\nu (T^A)_{\alpha \beta}\delta_{i j}$, where the
$T^A$ are the color Gell-Mann matrices. Notice that gluons in the CSC phase  acquire mass via the Anderson-Higgs mechanism and therefore the interaction between quarks should be properly described by the contact interaction above.

We want to use the non-renormalizable NJL model to describe the low energy behavior of the system, {\it i.e.}  the physics of quarks close to the Fermi surface.   In order to include all the pertinent quark fields and to avoid divergences we regulate the ultraviolet behavior of the loop integrals introducing a hard cutoff $\Lambda$,
which restricts the momentum integrals to a shell around the Fermi
sphere: $\mu-\Lambda<|\vec{p}|<\mu+\Lambda$. For this scheme to work we must chose   $\Lambda\ll\mu$ and at the same time it has to be that $\Lambda \gg \Delta_0$. If the relations above are not satisfied   
the theory will not be internally consistent.

In  the weak-coupling
approximation one finds that  $\Delta_0\ll\mu$ and it is possible to
choose $\Lambda$ such that $ \Delta_0\ll \Lambda \ll \mu$.
In the strong coupling region model calculations suggest that $\Delta_0$ should lie between
$10\,$MeV and $100\,$MeV~\cite{reviews}.  Therefore the above inequalities are still satisfied if say $\mu =300 - 400$ MeV and  $\Lambda \sim 200$ MeV.

The NJL model allows us to study the various color superconducting phases, including those with non-homogeneous condensates. However,  let us  consider for simplicity homogeneous phases.  
In the mean field approximation  the interaction Lagrangian
(\ref{interactionLagrangian}) takes on the form
\begin{equation}
{\cal L}_{\rm interaction}= \frac{1}{2}\bar{\psi}\Delta_0\bar{\psi}^T +
\frac{1}{2} \psi^T\bar{\Delta}_0\psi - \frac{3}{8}\,G\, {\rm
tr}\bigl(\Gamma^T\langle\bar{\psi}^T\bar{\psi}\rangle
  \Gamma\bigr)\langle\psi\psi^T\rangle,
\label{meanfieldapprox}
\end{equation}
where, $ {\rm tr}$ represents the trace over color, flavor and Dirac
indices  and where the gap parameter is related to the diquark condensate
by the relations,
\begin{equation}
\Delta_0 = \frac{3}{4}\,G\,\Gamma\langle\psi\psi^T\rangle\Gamma^T 
\;\;{\mbox{ and}}\;\;\;\;
\bar{\Delta}_0 = \gamma^0\Delta^{\dagger}({\bf r})\gamma^0 \label{deltaislambdacondensate}\;.
\end{equation}
In order to understand the physical meaning of the gap parameter  $\Delta_0$, we notice that 
the interaction term in Eq.~(\ref{meanfieldapprox}) takes on the form of a   Majorana mass for the quark fields. This mass, however, is not diagonal in flavor and color indices, and  to obtain a Lagrangian describing a system of non-interacting quasiparticles, the quark fields have to be properly rotated. The decoupled modes  will describe the quasiparticle excitation of the system with a dispersion law having  a gap proportional to $\Delta_0$. Notice that the fact that the Majorana mass is not diagonal in color and flavor indices, means that color  and flavor symmetries are broken, as it should be.

\section{Lower  densities}
To describe quark matter as may exist in the cores of compact stars,
we need to consider quark number chemical potentials $\mu$ of order
$500$~MeV at most (corresponding to $\mu_B=3\mu$ of order $1500$~MeV).
For these values of the chemical potentials  many problems arise.
Clearly the QCD coupling is not small and perturbation theory cannot be applied.  
Moreover, the relevant range of $\mu$ is low enough that the QCD  strange quark
mass, $M_s$, lying somewhere between its current mass of order
$100$~MeV and its vacuum constituent mass of order $500$~MeV, cannot
be neglected. Furthermore, bulk matter, if present in neutron stars,
must be in weak equilibrium and must be electrically and color neutral. 

The combined effect of  strange quark mass, of weak equilibrium and of color and electrical neutrality is to induce a mismatch $\delta\mu \sim M_s^2/\mu$ between the  Fermi momenta. 
For simplicity  suppose we start by setting the gauge chemical potentials
to  zero. Then, in weak equilibrium a nonzero $M_s$ implies
that there are fewer $s$ quarks in the system than $u$ and $d$ quarks, and
hence the system is positively charged. To restore electrical
neutrality, a positive $\mu_e$ is required which tends to reduce the
number of up quarks relative to the number of down and strange
quarks. More in detail, in the absence of pairing, color neutrality is obtained simply with
$\mu_3=\mu_8=0$, whereas  weak equilibrium implies that
\begin{equation} \mu_u =  \mu - \frac{2}3 \mu_e\,,\hspace{0.5cm}   \mu_d = \mu_s=  \mu + \frac{1}3 \mu_e\,.\end{equation} Electrical neutrality requires that the number densities satisfy the equation 
\begin{equation} \label{eneutrality}
\frac{2}3 N_u - \frac{1}3 N_d - \frac{1}3 N_s - N_e=0 \,,\end{equation} and Fermi momenta are given by 
\begin{equation}
p_u^F = \mu_u, \hspace{0.5cm} p_d^F = \mu_d,  \hspace{0.5cm}  p_s^F = \sqrt{\mu_s^2 -M_s^2}, \hspace{0.5cm} p_e^F = \mu_e\,.\end{equation}
 Expanding the Fermi
momentum of the strange quark in  $M_s^2/\mu^2$ we obtain
\begin{equation}
p_F^s=\sqrt{\left(\mu+\frac{\mu_e}{3}\right)^2-M_s^2}\approx
\left(\mu+\frac{\mu_e}{3}\right)-\frac{M_s^2}{2\mu}+{{O}}\left(\frac{M_s^4}{\mu^3}\right)\, \label{pF1},
\end{equation}
and  to this order electric neutrality, Eq.~(\ref{eneutrality}), requires that
$\mu_e=\frac{M_s^2}{4\mu}$, yielding
\begin{equation}
p_F^d  = p_F^u+\frac{M_s^2}{4\mu},  \hspace{0.5cm}
p_F^u = \mu-\frac{M_s^2}{6\mu},  \hspace{0.5cm}
p_F^s = p_F^u-\frac{M_s^2}{4\mu},  \hspace{0.5cm}p_F^e = \frac{M_s^2}{4\mu}\, .\label{pF2}
\end{equation}

A mismatch between the Fermi momenta  disfavors cross-species  pairing for the following reason. 
In order to form Cooper pairs,  gaining a  free energy $\sim \mu^2 \Delta_0^2$, one needs a pair of fermions with  zero total momentum. Therefore,  quark excitations far from the Fermi sphere have to be created, but this  process has a free energy cost $\sim \mu^2 \delta\mu^2$. Comparing the free energy of  the  CFL phase with the free energy of the  unpaired phase one finds that for 
$\frac{M_s^2}{4 \mu} < \Delta_0$ the CFL phase is energetically favored. This result is quite general and it always happens that  for mismatches smaller than  a critical mismatch  the homogeneous superconducting phase is favored over the unpaired phase. This critical mismatch is  generally called the Chandrasekhar-Clogston limit~\cite{Clogston:1962}. However, even for mismatches between Fermi momenta below this  limit, it is  not guaranteed that the homogenous superconducting phase is stable. For CFL matter,   when 
$\frac{M_s^2}{2 \mu} > \Delta_0$ gapless modes appear and the system is "chromo-magnetically unstable"~\cite{Casalbuoni:2004tb}, meaning that the masses of some gluons become imaginary and the energy can be lowered by the formation of counter-propagating currents~\cite{Huang:2004bg}. 
Let me emphasize  that the strongest constrain on the maximum chemical potential stress sustainable by the CFL phase does not come from the Chandrasekhar-Clogston limit, but comes from the "chromo-magnetic" instability.  

Suppose for simplicity that the strange quark mass does not depend on density. Then, if the   gap parameter is sufficiently large, say of order $100$ MeV, and the strange quark mass is sufficiently small, say about $200$ MeV, one has that  $\frac{M_s^2}{2  \Delta_0} \sim 200 $ MeV, and  the CFL phase would be the only color superconducting phase. This means that in the phase diagram on the right hand side of Fig.\ref{phasediagram}, the only CSC phase appearing on the right of $\mu_c$ would be the CFL phase. However, we do not have detailed control on the ratio  $\frac{M_s^2}{2  \Delta_0}$ and cannot exclude {\it a priori} the possibility that   $\frac{M_s^2}{2  \Delta_0} $ is large. As an example, for $M_s = 300$ MeV and $\Delta_0 \sim 50 $ MeV we have that $\frac{M_s^2}{2  \Delta_0} = 900$ MeV and such a large value completely  excludes the possibility that CFL quark matter can be found in the interior of compact stars.  In general, imagine we begin at asymptotically high densities and reducing the density  
the CFL pairing is disrupted before color superconducting quark matter is superseded by baryonic
matter. Then, the CFL phase must be replaced by some other superconducting phase of quark matter which has less symmetric pairing.

It is still an open challenge to determine the  favored color superconducting phase for  large values of $\frac{M_s^2}{2  \Delta_0}$. In  Refs.~\cite{Casalbuoni:2005zp,Mannarelli:2006fy,Rajagopal:2006ig}
the possibility that the superconducting phase has a crystalline structure was explored. Crystalline color
superconductivity --- the QCD analogue of a form of non-BCS pairing
first considered by Larkin, Ovchinnikov, Fulde and Ferrell
(LOFF)~\cite{LOFF} --- is an attractive candidate phase in the
intermediate density regime
\cite{Alford:2000ze},
because it allows quarks living on split Fermi surfaces to pair with
each other. It does so by allowing Cooper pairs with nonzero total
momentum. 
In position space, this corresponds to condensates  which are modulated periodically in space
and therefore spontaneously break space translation invariance.
This phase seems to be free from magnetic instability~\cite{Ciminale:2006sm}.

The free energies of various periodic structures are compared in the Ginzburg-Landau approximation in Ref.~\cite{Rajagopal:2006ig}. Unfortunately, this approximation is not under quantitative control, because the control parameter for this expansion, $\Delta/(M_s^2/8\mu)$,
is about $1/2$, meaning that the expansion is pushed to or beyond the
boundary of its regime of quantitative validity.  For
analysis which go beyond the Ginzburg-Landau approximation
see~\cite{Mannarelli:2006fy,Nickel:2008ng}.

Other possible patterns of three-flavor and three-color pairing are reported in Refs.~\cite{Kryjevski:2005qq,Gerhold:2006dt} where the  instability of the CFL state  is resolved by the formation of an inhomogeneous meson condensate. 
A different possibility is the  2SC phase where  only two flavors and two colors of quarks pair, ~\cite{Alford:1997zt,Rapp:1997zu}. However,  also in this case local charge neutrality and weak equilibrium induce a mismatch between the Fermi momenta of quarks of different flavors~\cite{Shovkovy:2003uu} and the corresponding 2SC gapless phase turns out to be unstable~\cite{Huang:2004bg}. In this case the  chromo-magnetic instability may be cured if the gluonic dynamics leads to the spontaneous breakdown of the color gauge symmetry, (along with $U(1)_{\rm e.m.}$ and the rotational $SO(3)$) in the so-called gluonic phase~\cite{Gorbar:2005rx}. Finally, the mismatch between Fermi momenta may be so large that  only quarks of the same flavor can pair. In this case the fermionic gap would be at most of the order of $1 $ MeV, much smaller than the gap that characterizes phases with cross-pairing.  For a detailed analysis of the various CSC phases that may arise in this case  see Refs.~\cite{Iwasaki:1994ij}.

\end{document}